\documentclass[aps,english,showpacs,twocolumn,pra]{revtex4-1}
\usepackage{amsfonts}
\usepackage{amssymb}
\usepackage{amsmath}
\usepackage{graphicx}
\usepackage{epsfig}
\usepackage{color,xcolor}

\begin{document}

\title{Signature of edge states in resonant wave scattering}
\author{H. S. Xu}
\author{K. L. Zhang}
\author{L. Jin}
\email{jinliang@nankai.edu.cn}
\author{Z. Song}
\email{songtc@nankai.edu.cn}
\affiliation{School of Physics, Nankai University, Tianjin 300071, China}

\begin{abstract}
{Particle beam scattering is a conventional technique for detecting the
nature of matter. We studied the scattering problem of a cluster connected
to multiple leads. We established the connection between the eigenstate of
the topological scattering center and the transmission and reflection
amplitudes for the resonant scattering process. We discovered that as an
application, this approach enables the detection of the edge state in the
band gap for both Hermitian and non-Hermitian systems and the identification
of the topology of a system. We investigated two types of
Su-Schrieffer-Heeger chains as examples. In addition, we proposed a dynamic
scheme through an evanescently coupled-waveguide array to detect the edge
state on the basis of measured transmission intensity. Numerical simulation
revealed that pattern visibility can be the signature of the edge states.}
\end{abstract}

\maketitle

\section{Introduction}

Studies on the integer quantum Hall effect have indicated that edge modes
exist in the topological phase with nonzero Chern numbers \cite%
{HalperinPRB82}. According to band theory, the gap closing between filled
and empty bands will motivate the existence of edge states \cite{CooperRMP19}%
. The result is a robust feature for any form of boundary, called the
bulk-boundary correspondence \cite{BansilRMP16}. The only restriction is
that, in cases where the topological invariant relies on an underlying
symmetry, this symmetry must be preserved also in the boundary region. A
typical example is the Su-Schrieffer-Heeger (SSH) model, in which chiral
symmetry is crucial \cite{AsbothBook16}. Another key feature of the edge
state is that its wavefunction exponentially localizes on the boundary. The
almost-zero-energy eigenstates of SSH model are odd and even superpositions
of states localized exponentially on the left and right edges. This is a
consequence of the exponentially small overlap between the left and right
edge states.

Photonic crystal is an excellent platform for the study of topological
physics \cite{OzawaRMP19}. Topological photonic devices have been realized
in microwave-scale magnetic photonic crystals \cite%
{HaldanePRL08,RaghuPRA78,ZWangPRL08,ZWangNAT09} and meta-atom structures
\cite{WJChenNAT14}, function at optical and infrared frequencies in
waveguide lattices \cite{RechtsmanNAT13} and resonator lattices \cite%
{HafeziNAT11,HafeziNAT13}. Theoretical proposals based on modulated photonic
crystal resonances \cite{KFangNAT12}, circuit quantum electrodynamics
systems \cite{KochPRA10,PetrescuPRA12}, and metamaterial photonic crystals
\cite{KhanikaevNAT13} have also been developed. The key feature of these
devices is the existence of topologically protected electromagnetic edge
states.

A topological phase transition (TPT), which is caused by changes in the
topology of the bulk band structure, differs considerably from common phase
transitions such as the melting of a solid, which are characterized by
broken symmetries and sharp anomalies in thermodynamic properties \cite%
{BansilRMP16}. Numerous theoretical studies have demonstrated that TPTs can
be induced by tuning the band structure through chemical substitution,
strain, or pressure, or via electron correlation effects \cite%
{SatoPRB09,PesinNAT10,XWanPRB11,WrayNAT11,SYXuSCI11,LWuNAT13}; and can be
induced via laser or microwave pumping to produce a nonequilibrium
topological state or Floquet topological insulator \cite%
{KitagawaPRB10,ZGuPRL11,LindnerNAT11,DoraPRL12,KatanPRL13,KunduPRL13,YHWangSCI13,PiskunowPRB14,RWangEPL14}%
.

Topological states are characterized by topological invariant \cite%
{BernevigBook13}. In fermionic systems, conductance measurements reveal
these integer invariants. However, direct measurement of these integers is
non-trivial in bosonic systems, mainly because the concept of conductance is
not well defined \cite{OzawaPRL14,HafeziPRL14}. Whereas these integers have
been measured in one-dimensional (1D) bosonic systems \cite{WHuPRX15,AtalaNAT14,DucaSCI15},
the two-dimensional (2D) bosonic case has been realized in atomic lattices \cite%
{AidelsburgerSCI15} and photonic system \cite{MittalNAT16}.

Recently, topological concepts have been applied to scattering. Topological
Fano resonance is immune to impurities, although it remains sensitive to
system parameters \cite{NejadPRL19}. A receiver protector was proposed and
demonstrated employing a topological interface state of the SSH lattice \cite%
{ReisnerPRA20}. Scattering methods to measure topological invariant have
also been developed. The winding number of scattering matrix eigenvalues
determines the number of edge states and topological invariants \cite%
{MeidanPRB11,FulgaPRB12,RudnerPRX13,PasekPRB14,LJinPRA17,HCWuPRB19}. The
relationships among surface scattering properties, bulk band properties, and
the formation of interface states for a 1D centrosymmetric
photonic crystal have been revealed \cite{MXiaoPRX14}. In addition, the
phase of the reflection coefficient can be used to measure the topological
indexes of a photonic system \cite{PoshakinskiyPRA15,ArkinstallPRB17}.

In this paper, we investigated the relationship between the transmission
coefficients of multiple output channels scattering in the resonance process
and the edge state of the topological scattering center, which can be used
to intuitively detect the edge state. The proposed method is insensitive to
system perturbation. We also developed a method of measuring real
eigenenergy by determining whether perfect transmission occurs in a two-lead
scattering system. We demonstrated the effectiveness of this method by
applying it to detect an edge state in an SSH model and designed a
corresponding experimental platform through a system of waveguide arrays.
Through numerical simulation, we show that the proposed scattering system
can be used to distinguish different phase regions from the visibility and
reflection. We also show that our proposed scattering formalism can be used
to detect the eigenfunctions of non-Hermitian systems with real energy
spectra.

The remainder of this paper is organized as follows. In Sec.~\ref{II}, we
design a multi-transmission channels scattering system and find the
relationship between transmission coefficients and the eigenfunction of the
topological scattering center. In Sec.~\ref{III}, we take an SSH chain as
scattering center and proved that our conclusion is applicable to detect the
edge state. In Sec.~\ref{IV}, we provide the scheme to experimentally detect
the edge state via a system of waveguide arrays, and investigate the
visibility and reflection to identify different phase regions by numerical
simulations. In Sec. \ref{V}, we consider a non-Hermitian SSH chain by
adding the imaginary potential $\pm i\gamma $, and we show the transmission
probability in each leads is conformed to its eigenfunction, which is a
sinusoidal function. Finally, we summarize the results and conclude in Sec. %
\ref{VI}.

\section{General formalism}

\begin{figure}[tb]
\includegraphics[bb=-15 300 255 425, width=8.8 cm, clip]{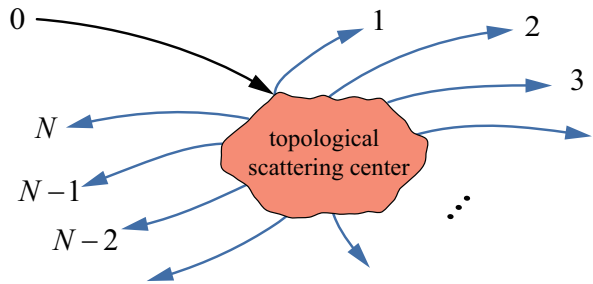}
\caption{Illustration of multi-transmission channel scattering system for detecting in-gap eigenstates. The orange area indicates $N$-site topological scattering center $H_{\mathrm{c}}$, and each site $\left\vert l\right\rangle _{\mathrm{c}}$ in the scattering center is connected to
$l$-th transmission chain ($l\in \lbrack 1,N]$). The incident and
transmission chains are both tight-binding chains, and coupling strength $\left\vert J\right\vert \ll 1$. }
\label{fig1}
\end{figure}

\label{II}

In this section, we provide the general scattering formalism for the
detection of in-gap eigenstates. The transmission coefficients for the
on-resonance input when all output channels are weakly connected to the
scattering center approximately indicate the eigenstates of the scattering
center. Figure~\ref{fig1} presents the structure of the proposed scattering
system. The Hamiltonian of the system reads%
\begin{equation}
H=H_{\mathrm{c}}+H_{\mathrm{in}}+H_{\mathrm{out}}+H_{\mathrm{jnt}},
\label{E1}
\end{equation}%
where $H_{\mathrm{c}}$ represents the topological scattering center, namely
an $N$-site lattice and each site is attached to one output channel. The
channels are semi-infinite tight-binding chains with uniform coupling
strength $J$. We only considered cases with single input channels. The
Hamiltonian of the input channel is%
\begin{equation}
H_{\mathrm{in}}=\sum\limits_{j=1}^{\infty }(J\left\vert -j\right\rangle
_{0}\left\langle -j-1\right\vert _{0}+\mathrm{H.c.}+\mu \left\vert
-j\right\rangle _{0}\left\langle -j\right\vert _{0}),  \label{E2}
\end{equation}%
where $\left\vert j\right\rangle _{0}$ is the single-particle basis of the
input leads at site $j$. $H_{\mathrm{out}}$ represents transmission channels
and reads%
\begin{equation}
H_{\mathrm{out}}=\sum\limits_{l=1}^{N}\sum\limits_{j=1}^{\infty
}(J\left\vert j\right\rangle _{l}\left\langle j+1\right\vert _{l}+\mathrm{%
H.c.}+\mu \left\vert j\right\rangle _{l}\left\langle j\right\vert _{l}),
\label{E3}
\end{equation}%
where $\left\vert j\right\rangle _{l}$ is the basis at site $j$ in the $l$%
-th output lead. The joint Hamiltonian is%
\begin{equation}
H_{\mathrm{jnt}}=J\left\vert -1\right\rangle _{0}\left\langle 1\right\vert _{%
\mathrm{c}}+J\sum\limits_{l=1}^{N}\left\vert 1\right\rangle _{l}\left\langle
l\right\vert _{\mathrm{c}}+\mathrm{H.c.},  \label{E4}
\end{equation}%
where $\left\vert l\right\rangle _{\mathrm{c}}$ denotes the basis of
scattering center $H_{\mathrm{c}}$ connected to the $l$-th transmission
channel. Here, $J$\ denotes the hopping strength and $\mu $ denotes the
chemical potential for each lead. The entire scattering system consists of
the scattering center and the input and output leads. Energy exchange
between the leads and the scattering center is enabled by weak couplings.

The wavefunction for the input lead is represented by $\psi _{0}^{k}$, and
the $l$-th output lead is represented by $\psi _{l}^{k}$, where $k$ is the
dimensionless wave vector. The incoming plane wave is reflected and
transmitted by the scattering center, and only outgoing plane waves are
present in all output leads. The wavefunctions in the leads are given by
\begin{equation}
\left\{
\begin{array}{ll}
\psi _{0}^{k}\left( j\right) =e^{ikj}+re^{-ikj}, & j<0; \\
\psi _{l}^{k}\left( j\right) =t_{l}e^{ikj}, & j>0,%
\end{array}%
\right.  \label{E5}
\end{equation}%
where $r$ and $t_{l}$ are the reflection and transmission coefficients for
the input and $l$-th\ output leads, respectively. The corresponding
wavefunction in the scattering center is represented by $\psi _{\mathrm{c}%
}^{k}\left( j\right) $ ($j\in \lbrack 1,N]$). According to the continuity
condition of the wavefunction of a discrete quantum system, $\psi _{\mathrm{c%
}}^{k}\left( 1\right) =1+r$ and $\psi _{\mathrm{c}}^{k}\left( l\right)
=t_{l} $ ($l\in \lbrack 1,N]$) because of Eq.~(\ref{E5}). Thus,
\begin{equation}
1+r=t_{1}.  \label{E6}
\end{equation}%
The leads are tight-binding chains with a uniform hopping strength and
identical chemical potential. Dispersion relation $E_{k}=2J\cos k+\mu $ can
be obtained from the Schr\"{o}dinger equations for lead Hamiltonians $H_{%
\mathrm{in}}$ and $H_{\mathrm{out}}$.

Accordingly, the Schr\"{o}dinger equation for scattering center $H_{\mathrm{c%
}}$ is
\begin{equation}
(H_{\mathrm{c}}-E_{k})\left(
\begin{array}{c}
t_{1} \\
t_{2} \\
\vdots \\
t_{N-1} \\
t_{N}%
\end{array}%
\right) =-J\left(
\begin{array}{c}
(e^{-ik}+re^{ik})+t_{1}e^{ik} \\
t_{2}e^{ik} \\
\vdots \\
t_{N-1}e^{ik} \\
t_{N}e^{ik}%
\end{array}%
\right) ,  \label{E7}
\end{equation}%
which connects the eigenproblem of $H_{\mathrm{c}}$\ to the scattering
problem. For a given explicit form $H_{\mathrm{c}}$, $r$ and $t_{l}$ can be
obtained. This analysis is not restricted to a Hermitian $H_{\mathrm{c}}$.
We focused on obtaining a solution with $k\in \left[ 0,\pi \right] $\ and
real $\mu $, which requires that the corresponding eigenenergy be real.

Equation~(\ref{E7}) reduces to the secular equation of $H_{\mathrm{c}}$
\begin{equation}
(H_{\mathrm{c}}-E_{k})\left(
\begin{array}{ccccc}
t_{1}, & t_{2}, & \cdots , & t_{N-1}, & t_{N}%
\end{array}%
\right) ^{T}=0,  \label{E8}
\end{equation}
in the limit of zero $J$. When the input plane wave is in resonance with the
eigenenergy of $H_{\mathrm{c}}$, transmission coefficient $t_{l}$ approaches
the eigenfunction of $H_{\mathrm{c}}$; otherwise, if it is not a resonant
scattering process, all transmission coefficients $t_{l}$ vanish, and the
reflection coefficient $r$ becomes $-1$ in accordance with Eq.~(\ref{E6}).
Therefore, we established the relationship between the scattering
coefficients and the eigenfunction of the scattering center.

This method is particularly suitable for detecting edge states in
topological systems. The in-gap eigenstate exponentially localizes on the
boundary, and the wavefunction in the first unit cell is the largest. If the
input channel is connected to the site at which the wavefunction is the
largest, then the transmission probability of the first output channel will
be much larger than the other output channels, and the reflection
probability will be the lowest according to Eq.~(\ref{E6}). The reflection
should be as small as possible to observe large transmissions. In addition,
this helps distinguish nonresonant cases, which have total reflection and
zero transmission. In the following, we used 1D topological
models as examples to demonstrate edge state detection.

\section{Dynamic detection of edge state}

\label{III}

\begin{figure}[tb]
\includegraphics[bb=0 0 420 325, width=8.8 cm, clip]{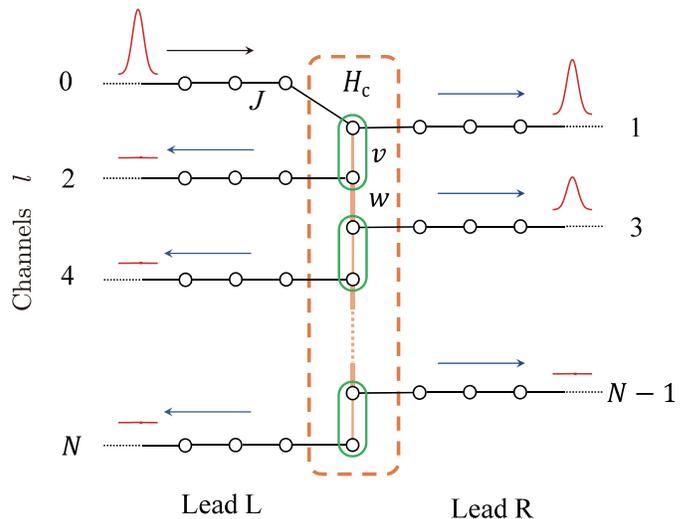}
\caption{A SSH model scattering center in dashed orange box with
intracell hopping $v$ (thin lines) and intercell hopping $w$ (thick lines).
Even (odd) channels are placed on the left (right). In the left channels, lead site $j \in [-\infty,-1]$, and in the channel on the right, lead site $j \in [1,\infty]$.
The transmission probability in each lead is a geometric
sequence for topological states ($v<w$) and to be zero for trivial states ($v>w $). }
\label{fig2}
\end{figure}

\begin{figure*}[tb]
\includegraphics[bb=20 0 1688 1110, width=17.8 cm, clip]{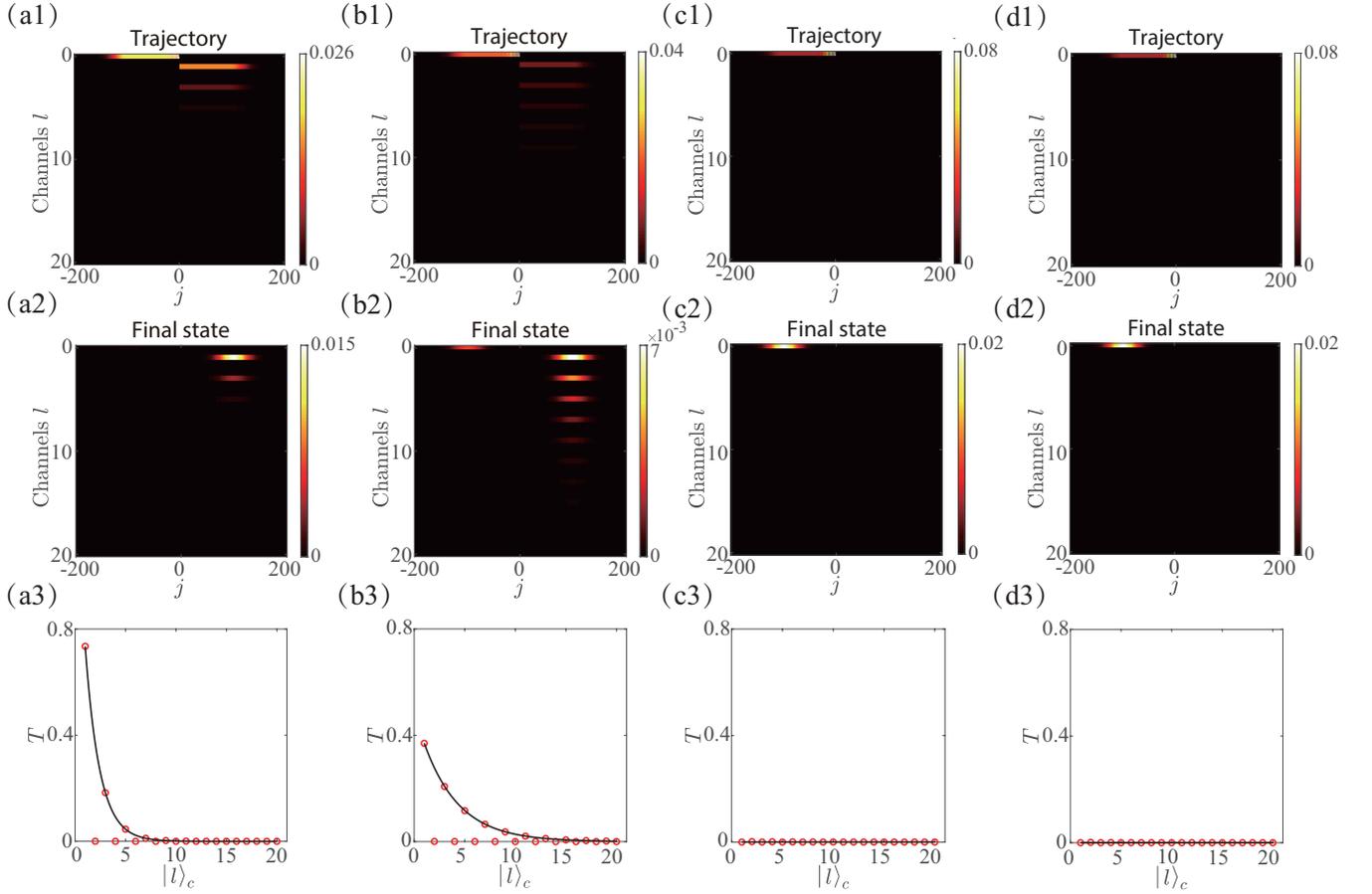}
\caption{Profile of trajectory and final state for the injected Gaussian
wave packet with several typical sets of scattering center parameters. The
system consists of a 40-site scattering center (20 unit cells) and 41
200-site tight-bounding chains. The plots present only the first 20
channels, and the transmission probabilities of the last 20 channels are
almost zero. A Gaussian wave packet is initially centered at site $N_{
\mathrm{c}}=-100 $. The wave vector for the Gaussian wave packet is $k_{
\mathrm{c}}=\protect\pi/2, \protect\sigma=20 $, and $\protect\mu=0$ to
ensure $E_k=0$ and resonance with the in-gap eigenmode. We set coupling
strength $J=-0.1$. (a), (b), (c), and (d) correspond to $w=4$ and $v=2 $, $%
v=3$, $v=5$ and $v=6$, respectively. (1) and (2) correspond to the
trajectory and final state for each parameter. The overall structure is
similar to that in Fig.~\protect\ref{fig2}. The vertical axis represents the
channels number, and the horizontal axis represents the left and right leads
with 200 sites. (3) is a comparison of transmission probabilities (red empty
circle) and the eigenfunction (black line for odd sites), and the
eigenfunction has been renormalized using the Eq.~(\protect\ref{E14}). The
transmission probability is a geometric series for the topological case and
almost zero for the trivial case.}
\label{fig3}
\end{figure*}

The previous section demonstrated that the eigenfunction can be obtained by
using transmission probabilities with eigenenergy known. In this section, we
consider an SSH chain as a scattering center to demonstrate the proposed
method. The SSH model is a prototypical 1D topological model
with typical edge state features and robustness to disorder. In addition,
The SSH model is also the core of numerous other topological models,
including the SSH ladder \cite{KLZhangPRB19,HCWuPRB20,HCWuPRB21}, the
second-order topological insulator \cite%
{BYXiePRB18,BomantaraPRB19,FukuiPRB19}, and the SSH-Hubbard model \cite%
{LibertoPRA16,LibertoEPJ17}. Therefore, we used the SSH model as an example
without loss of generality. The Hamiltonian $H_{\mathrm{c}}$ is
\begin{equation}
H_{\mathrm{c}}=\sum\limits_{m=1}^{N/2}v\left\vert 2m-1\right\rangle _{%
\mathrm{c}}\left\langle 2m\right\vert _{\mathrm{c}}+\sum%
\limits_{m=1}^{N/2-1}w\left\vert 2m\right\rangle _{\mathrm{c}}\left\langle
2m+1\right\vert _{\mathrm{c}}+\mathrm{H.c.},  \label{E9}
\end{equation}%
and Fig.~\ref{fig2} presents the scattering system. We only focused on cases
in which $\mu =0$ and $k=\pi /2$. At large $N$ limits, the system occupies a
topological phase when $v<w$, which has an in-gap zero-energy edge state in
the form of
\begin{equation}
\left\vert \phi _{0}\right\rangle
=(1-q^{2})^{1/2}\sum_{j=1}^{N/2}(-q)^{j-1}\left\vert 2j-1\right\rangle _{%
\mathrm{c}},  \label{E10}
\end{equation}%
($q=v/w$), and a trivial phase when $v>w$, for which edge state is absent.
In Appendix A, we used another scattering system to demonstrate the zero
mode existing in a topological phase and vanishing in a trivial phase. In
the next, we will show that the wavefunction of edge states can be detected
using the amplitudes of outgoing waves in a topological phase, and the
amplitudes are zeros in a trivial phase.

For numerical simulation, the initial state was used as the Gaussian wave
packet, given by%
\begin{equation}
\left\vert \varphi (0)\right\rangle =\Omega
_{0}^{-1/2}\sum\limits_{j}e^{-(j-N_{\mathrm{c}})^{2}/2\sigma ^{2}}e^{ik_{%
\mathrm{c}}j}\left\vert j\right\rangle _{0},  \label{E11}
\end{equation}%
where $\Omega _{0}=\sum\limits_{j}e^{-(j-N_{\mathrm{c}})^{2}/\sigma ^{2}}$
is the normalization factor, $k_{\mathrm{c}}$ is the central wave vector of
the Gaussian wave packet, and the full width at half maximum of the
intensity of Gaussian wave packet is $2\sqrt{\ln 2}\sigma $. For the
simulation of plane waves, the width of the incoming wave packet must be
large; otherwise, the dynamics are mixed with the dynamics near $k_{c}$.
Initially, the Gaussian wave packet was centered at site $N_{\mathrm{c}}$,
within the input lead. The evolved state is computed as follows
\begin{equation}
\left\vert \varphi (t)\right\rangle =e^{-iHt}\left\vert \varphi
(0)\right\rangle ,  \label{E12}
\end{equation}%
which yields the probability summations in each channel
\begin{equation}
p_{l}=\left\{
\begin{array}{cc}
\sum\limits_{j=1}\left\vert \langle \varphi (T)\left\vert +j\right\rangle
_{l}\right\vert ^{2}, & \text{odd }l \\
\sum\limits_{j=1}\left\vert \langle \varphi (T)\left\vert -j\right\rangle
_{l}\right\vert ^{2}, & \text{even }l%
\end{array}%
\right.  \label{E13}
\end{equation}
($l\in \lbrack 1,N]$), where $T$ is the time at which the scattering process
finishes. Relations $p_{0}=\left\vert r\right\vert ^{2}$\ and $%
p_{l}=\left\vert t_{l}\right\vert ^{2}$\ ($l>0$) should be observed. Because
the entire scattering system is Hermitian, it satisfies the conservation of
probability. Therefore, for the input Gaussian wavepacket,
\begin{equation}
p_{l}=\left\{
\begin{array}{ll}
\lbrack 1-q^{2}/\left( 2-q^{2}\right) ]^{2}q^{l-1}, & \text{odd }l \\
\lbrack q^{2}/\left( 2-q^{2}\right) ]^{2}, & l=0 \\
0, & \text{even }l\neq 0%
\end{array}%
\right.  \label{E14}
\end{equation}%
should be observed for $v<w$, whereas%
\begin{equation}
p_{l}=\left\{
\begin{array}{cc}
1, & l=0 \\
0, & l>0%
\end{array}%
\right.  \label{E15}
\end{equation}%
should be observed for $v>w$. Appendix B presents the derivation of the
analytical results. To demonstrate our prediction, we plotted the numerical
results for several typical sets of parameters in Fig.~\ref{fig3}. A
comparison between the analytical and numerical results revealed that the
wave packet can be used to demonstrate the conclusion on the plane wave. Our
conclusion is invalid near topological phase transition point $v=w$ because
the zero mode is absent in small finite SSH chains. The outputs from the
left (right) channel in the numerical simulations yield the left (right)
edge state.

\section{Dynamic detection of phase region}

\label{IV} This section provides a scheme to experimentally detect edge
states through a system of waveguide arrays. Recently, numerous studies have
experimentally demonstrated \textrm{SSH} models in photonic systems \cite%
{LLuNAT14,OzawaRMP19}. It is based on the analogy between light propagating
through a photonic crystal and a tight-binding Hamiltonian. Topological
effects in some electronic systems can be observed in their photonic
counterparts \cite%
{RechtsmanPRL13,HafeziNAT13,PlotnikNAT14}. On a
photonic platform, a single-particle state can be amplified by a large
population of photons. This enables a high degree of control over the system
parameters.

\begin{figure}[tb]
\includegraphics[bb=25 180 585 607, width=8.8 cm, clip]{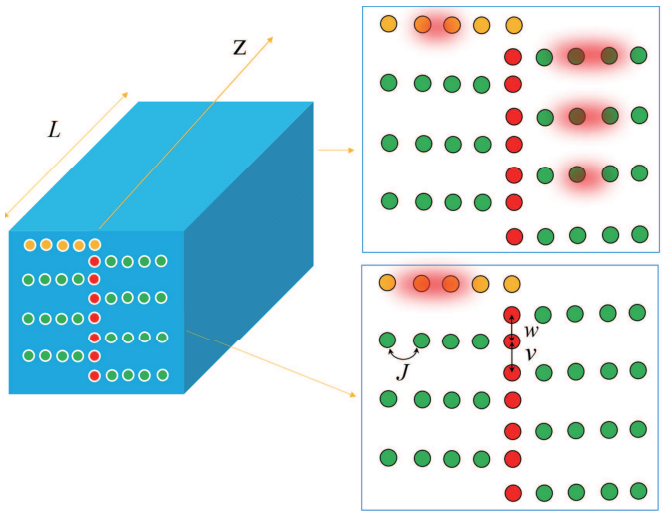}
\caption{Illustration of waveguide arrays for experimentally detecting the edge state.
The yellow dot represents the incident chain, the green dot represents the
transmission chain, and the red dot represents the SSH chain of the
scattering center. Each dot represents a waveguide of length $L$, which
should be designed to finish the scattering process in accordance with Eq.~(\protect\ref{E16}) and Eq.~(\protect\ref{E17}). Coupling strength $J$ in the
leads should be much smaller than coupling $v$ and $w$ in scattering center.
The bottom right and top right represent the initial and final states,
respectively. } \label{fig4}
\end{figure}

In the following we present a scheme to experimentally demonstrate of $p_{l}$
for $H_{\mathrm{c}}$ through a 2D array made of $\left( N+1\right) \times M$
equal straight waveguides, which are assumed to be weakly coupled. The
waveguides can be fabricated through the direct laser writing method, and
numerous 2D topological lattice systems have been implemented using this
technique \cite{NohNAT17,StutzerNAT18,NohNAT18}. Figure~\ref{fig4} presents
the geometry of the scattering system, which is uniform in the direction of
light propagation $z$. The input and each output lead are separated to
prevent the quantum tunneling effect. According to coupled-mode theory \cite%
{HausIEEE91,SHFanJOS03,JoannopoulosBook08}, light propagation can be
described by Schr\"{o}dinger-like equations, which are typically used to
mimic the dynamics of a tight-binding system.

For a coupled waveguide array with length $L$, the corresponding equations
are%
\begin{equation}
i\frac{\partial u(z)}{\partial z}=Hu(z),  \label{E16}
\end{equation}%
for $z\in \lbrack 0,L]$, with $u(z)=[u_{l,i}(z),u_{ \text{c},j}(z)]^{T}$
representing the vector, $u_{0,i}(z)$ ($i\in \left[ 1,N\right] $) denoting
the mode amplitude in the input waveguide, $u_{l,i}(z)$ ($l\in \left[ 1,M%
\right] ,i\in \left[ 1,N\right] $) denoting the mode amplitude in the output
waveguide, and $u_{\text{c},j}(z)$ ($j\in \left[ 1,M\right] $) denoting the
mode amplitude in the center waveguide. Accordingly, for a given initial
wave function $u(0)$,
\begin{equation}
u(L)=e^{-iHL}u(0).  \label{E17}
\end{equation}%
If $u(0)$ is used as a Gaussian function, as defined in Eq.~(\ref{E11}), and
light propagation distance $z$ is used as time $t$, then $u(L)$ corresponds
to the numerical results in Fig.~\ref{fig3}. This demonstrates that
detecting edge states is possible in arrays of coupled waveguides.

\begin{figure}[tb]
\includegraphics[bb=0 0 1220 615, width=8.8 cm, clip]{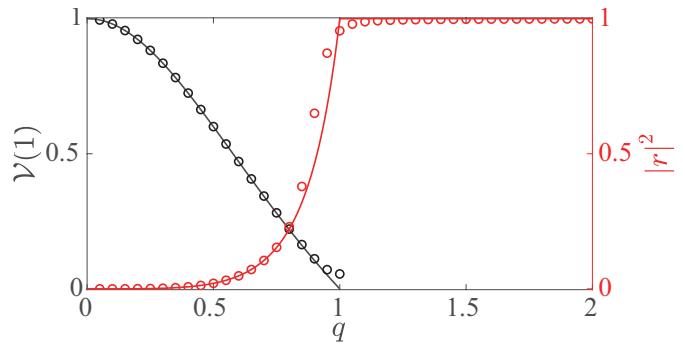}
\caption{$\mathcal{V}(1)$ (black) and $\left\vert r\right\vert ^{2}$
(red) as functions of $q$. The lines were obtained from the analytical
results of Eq.~(\protect\ref{E19}) and Eq.~(\protect\ref{E20}), and the empty
circles were obtained from the numerical simulations for an incident wavepacket. A non-analytical point was observed at $q=1$, which is the
topological phase transition point, and our theory adequately explained all $q$ values except those near the phase transition point. }
\label{fig5}
\end{figure}

We investigated the visibility and reflection to distinguish the
topologically nontrivial and trivial phases. According to the theoretical
analysis, the light intensity distribution in the output waveguide array
(see Fig.~\ref{fig3}) is $q$-dependent, thus obeying Eq.~(\ref{E14}). The
value of $q$ determines the visibility of two odd neighbor waveguides, which
is defined as%
\begin{equation}
\mathcal{V}(\eta )=\left\vert \frac{\sum_{i}\left\vert u_{2\eta
+1,i}(L)\right\vert ^{2}-\sum_{i}\left\vert u_{2\eta -1,i}(L)\right\vert ^{2}%
}{\sum_{i}\left\vert u_{2\eta +1,i}(L)\right\vert ^{2}+\sum_{i}\left\vert
u_{2\eta -1,i}(L)\right\vert ^{2}}\right\vert ,  \label{E18}
\end{equation}%
where $\eta \in \lbrack 1,M/2-1]$. The visibility in the region $q<1$ obeys%
\begin{equation}
\mathcal{V}(\eta )=\left( 1-q^{2}\right) /\left( 1+q^{2}\right) .
\label{E19}
\end{equation}%
Visibility $\mathcal{V}(\eta )$ in region of $q>1$ is not well defined
because all scattering chains are off-resonant with the energy levels of the
scattering center and then all transmission approach zero.

\begin{figure*}[tb]
\includegraphics[bb=20 0 1710 1110, width=17.8 cm, clip]{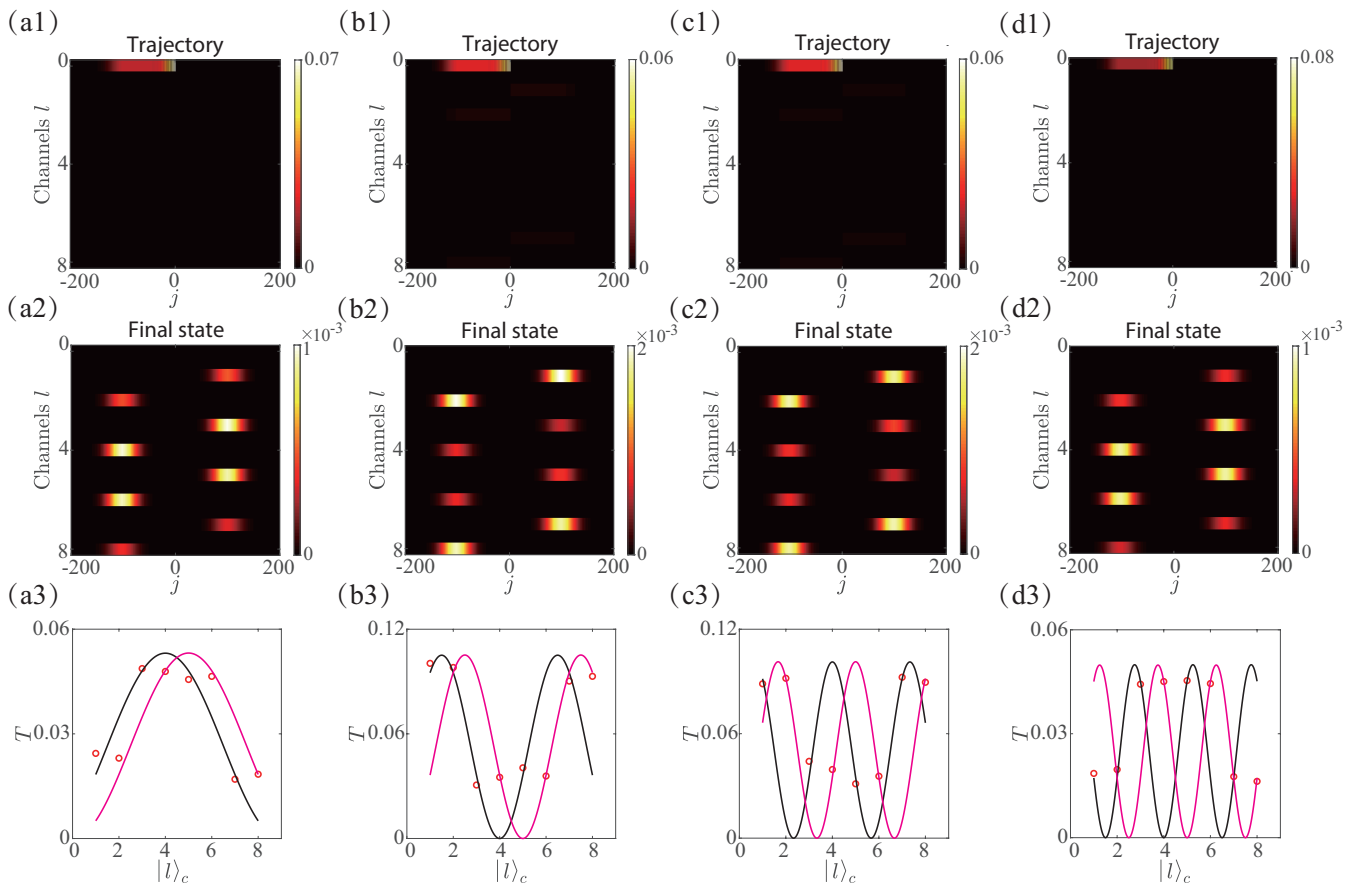}
\caption{Profile of trajectory and final state for the injected Gaussian
wave packet with several resonance energies. A Gaussian wave packet was
initially centered at the site $N_{\mathrm{c}}=-100 $ and $\protect\sigma=20$
. The parameters of scattering center were $v=40$, $w=2$, and $\protect\gamma%
=10$. The system consists of an eight-site scattering center and 9 200-site
tight-bounding chains. We set coupling strength $J=-0.1$. (a), (b), (c), and
(d) correspond to $n=1,2,3,$ and $4$, respectively. (1) and (2) correspond
to the trajectory and final state for each eigenenergy, respectively. (3) is
a comparison of the transmission probabilities (red empty circle) and the
eigenfunction (black line for odd sites and magenta for even sites), and the
eigenfunction has been renormalized to fit the transmission probabilities.
The transmission probability is a sinusoidal function with different
periods, determined by energy level $n$.}
\label{fig6}
\end{figure*}

Reflection $\left\vert r\right\vert ^{2}$ obeys
\begin{equation}
\left\vert r\right\vert ^{2}=\left\{
\begin{array}{ll}
q^{4}/\left( 2-q^{2}\right) ^{2}, & q<1 \\
1, & q>1%
\end{array}%
\right. ,  \label{E20}
\end{equation}%
which can be obtained from the analytical results. In Fig.~\ref{fig5} we
plotted $\mathcal{V}(1)$\ and $\left\vert r\right\vert ^{2}$\ as functions
of $q$, and the results indicate that there is a non-analytical point at $%
q=1 $. It seems that the observation of the reflection can be the witness of
the phase transition. However, the zero mode vanished near $q=1$, and our
theory was not applicable. In addition, by measuring the visibility in the
output waveguide array and the light intensity in the input waveguide array,
the corresponding $q$ of the topological region can be measured in the
experiment. To verify our prediction, we performed numerical simulation and
computed $\mathcal{V}(1)$ and $\left\vert r\right\vert ^{2}$ as function of $%
q$, for an incident wavepacket. Figure~\ref{fig5} presents the numerical
results with empty circles, in comparison with the analytical solutions
(solid line). As expected, the values of $\mathcal{V}(1)$ and $\left\vert
r\right\vert ^{2} $\ deviated from the analytical results as $q$\ approached
$1$. In other regions, the theoretical results perfectly fit the numerical
simulation results.

\section{Non-Hermitian SSH chain}

\label{V} In this section, we consider a non-Hermitian SSH chain as a
scattering center. The non-Hermiticity arose from opposite-site imaginary
potential $\pm i\gamma $. The scattering system is similar to that in Fig.~%
\ref{fig2}, with only the staggered gain and loss presented as the on-site
term of the scattering center. The Hamiltonian of the non-Hermitian
scattering center is%
\begin{equation}
\mathcal{H}_{\mathrm{c}}=H_{\mathrm{c}}+i\gamma
\sum\limits_{m=1}^{N}(-1)^{m}\left\vert m\right\rangle _{\mathrm{c}%
}\left\langle m\right\vert _{\mathrm{c}},  \label{E21}
\end{equation}%
where $H_{\mathrm{c}}$ is the Hamiltonian of the Hermitian SSH model in Eq.~(%
\ref{E9}). For this case, we used a strong dimerization limit, namely, $v\gg
w$, and it has been shown in the work \cite{KLZhangPRA18} that the
eigenstates are approximately given by
\begin{equation}
|\psi _{\kappa }^{\mathrm{c}}\rangle =\sqrt{\frac{2}{N/2+1}}%
\sum_{m=1}^{N/2}(-1)^{{m}}\sin \left( \kappa {m}%
\right) \frac{\left\vert 2m-1\right\rangle _{\mathrm{c}}-i\left\vert
2m\right\rangle _{\mathrm{c}}}{1-ie^{-i\varphi _{\kappa }}},  \label{E22}
\end{equation}
with eigenenergy given by
\begin{equation}
\varepsilon _{\kappa }=\sqrt{(v-w\cos \kappa )^{2}-\gamma ^{2}},  \label{E23}
\end{equation}%
where $\tan \varphi _{\kappa }=\gamma /\varepsilon _{\kappa }$\ and $\kappa =(n+1)\pi /(N/2+1)$, $n\in \lbrack 0,N/2-1]$. In
Appendix A, we verified the corresponding eigenenergy. We only considered
cases with real values for $\varepsilon _{\kappa }$. Spectrum $\varepsilon
_{\kappa }$\ was non-degenerate. The conclusion we obtained in the previous
section was applicable for finite $N$ in the limit of zero $J$. For an
incident plane wave with wave vector $k_{\mathrm{c}}=\pi /2$, chemical
potential $\mu $\ can be adjusted to $\varepsilon _{\kappa }$. The
transmission coefficients in each lead are a sinusoidal function given by
\begin{equation}
\left\vert t_{2m-1}\right\vert ^{2}=\left\vert t_{2m}\right\vert ^{2}\propto
\sin ^{2}\left( \kappa m\right) ,m\in \lbrack 1,N/2].
\label{E24}
\end{equation}%
Different resonant $\mu $ values result in different distribution of $%
\left\vert t_{l}\right\vert ^{2}$, which is determined by $n$.

To test this prediction, numerical simulations were performed in accordance
with the procedure in the previous section. The numerical results for
several typical sets of parameters are plotted in Fig.~\ref{fig6}. A
comparison between the analytical and numerical results revealed that our
scattering formalism for detecting eigenfunctions was applicable to
non-Hermitian topological models with real-value edge states.

\section{Discussion and Summary}

\label{VI}

The prototypical SSH model was employed to demonstrate the proposed method
of edge state detection. If Hamiltonian $H_{\mathrm{c}}$ is a high-order
degenerate topological system with degenerate edge states at the boundary of
the lattice, the resonant transmission in the scattering dynamics will yield
the linear superposition of the edge states for the high-order degenerate
edge states (more than one pair of degenerate left--right edge states). The
proposed method can still be used to identify nontrivial topology, where
transmissions in the leads exponentially decay in the topologically
nontrivial phase and approach zero for all leads in the trivial phase. The
method can also be applied to other topological models, including the
Rice-Mele model, the Aubry-Andr\'{e}-Harper model, and their
generalizations, and the edge states can be detected using resonant
transmission.

We developed a method to detect eigenstates in the band gap by using
multichannel scattering. The connection between the eigenstate of the
scattering center and the transmission and reflection amplitudes of the
resonant scattering process was established at the isolated energy level.
The method is applicable to both Hermitian and non-Hermitian scattering
centers if the isolated energy level is real and can be used to detect the
eigenstates of a system through resonant scattering. We demonstrated that
the edge state in an SSH chain can be detected using photon probability in
the output leads. We also proposed a scheme to demonstrate these results
through a system of waveguide arrays.

\section*{Acknowledgment}

This work was supported by National Natural Science Foundation of China
(Grant No.~11975128 and No. 11874225).

\section*{Appendix A: Determination of eigenenergy}

\label{Appendix A}

\begin{figure}[tb]
\includegraphics[bb=0 0 390 670, width=8.8 cm, clip]{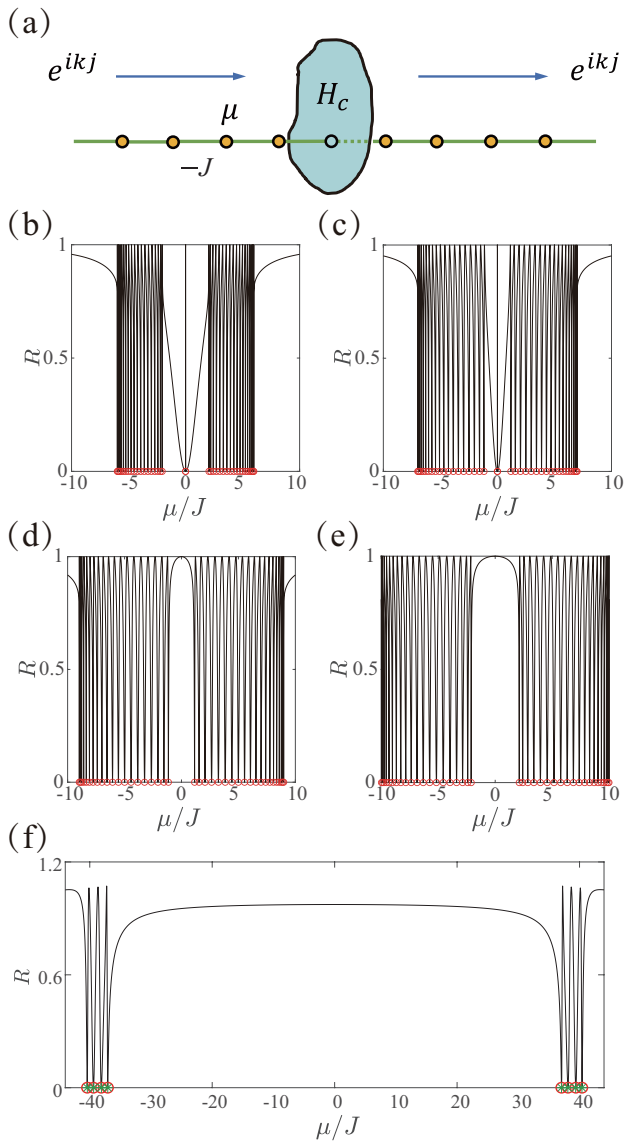} %
\caption{Schematic of the system for determining the eigenenergy of a
cluster and its corresponding numerical results. (a) Schematic of the
scattering system with two leads. (b)-(e) Numerical results of Hermitian SSH
model. The parameters for the scattering center were the same as those in
Fig.~\protect\ref{fig3}. (f) Numerical results of non-Hermitian SSH model.
The parameters for the scattering center were the same as those in
Fig.~\ref{fig6}. For all numerical results, coupling strength $J=1$, and
wave vector $k=\protect\pi/2$. The black lines are reflections with
different $\protect\mu$ values, and the red empty circles are the
corresponding eigenenergy of the scattering center. The green
asterisk in (f) represents the analytic eigenenergy in Eq.~(\ref{E23}).} \label{fig7}
\end{figure}

This appendix demonstrates the detection of the eigenenergy of a cluster
through the measurement of resonant transmission. We designed another
scattering system with two leads shown in Fig.~\ref{fig7}(a). The
Hamiltonian of the output lead in system $H$ reduces to%
\begin{equation}
H_{\mathrm{out}}=-J\sum_{j=1}^{\infty }(\left\vert j\right\rangle
\left\langle j+1\right\vert +\mathrm{H.c.})+\mu \sum_{\left\vert
j\right\vert =1}^{\infty }\left\vert j\right\rangle \left\langle
j\right\vert .  \label{E25}
\end{equation}%
The joint Hamiltonian is represented by
\begin{equation*}
H_{\mathrm{jnt}}=-J\left( \left\vert -1\right\rangle \left\langle \alpha
\right\vert +\left\vert 1\right\rangle \left\langle \alpha \right\vert
\right) +\mathrm{H.c.},
\end{equation*}%
where $\alpha $ represent an arbitrary site in cluster $H_{\mathrm{c}}$
connected to the left and right leads. $N$-site scattering center $H_{%
\mathrm{c}}$\ is%
\begin{equation}
H_{\mathrm{c}}=\sum_{n=1}^{N}\varepsilon _{n}\left\vert \phi
_{n}\right\rangle \left\langle \phi {_{n}}\right\vert .  \label{E26}
\end{equation}%
$H_{\mathrm{c}}$ is expressed by the eigenenergy representation. $\left\vert
\phi _{n}\right\rangle $ denotes the eigenstate of $H_{\mathrm{c}}$ with
energy $\varepsilon _{n}$ (if $H_{\mathrm{c}}$ is non-Hermitian,
$\left\langle \phi {_{n}}\right\vert $ in Eq. (\ref{E26}) should be replaced
by $\left\langle \varphi {_{n}}\right\vert $, where $\left\vert \varphi
_{n}\right\rangle $ denotes the eigenstate of $H_{\mathrm{c}}^{\dagger }$\
with energy $\varepsilon _{n}^{\ast }$). We considered a case in which $H_{%
\mathrm{c}}$\ has an isolated energy level at $\varepsilon _{q}$, satisfying
\begin{equation}
\varepsilon _{q}=-2J\cos k+\mu .  \label{E27}
\end{equation}%
State $|\psi _{k}\rangle $, calculated using
\begin{equation}
|\psi _{k}\rangle =\left\{
\begin{array}{ll}
e^{ikj}\left\vert j\right\rangle , & \left\vert j\right\vert \geqslant 1 \\
c\left\langle m\right\vert \phi _{q}\rangle \left\vert
m\right\rangle , & m\in \lbrack 1,N]%
\end{array}%
\right. ,  \label{E28}
\end{equation}%
is an eigenstate of $H$ with energy $\varepsilon _{q}$. Here $\left\{
\left\vert m\right\rangle \right\} $\ denotes the set of indexes for the
sites of the scattering center, and $c$ is a complex number,
calculated using $\left\langle \alpha \right\vert \phi _{q}\rangle =c^{-1}$,\ which implies that $\left\langle \alpha \right\vert
\phi _{q}\rangle $\ must be non-zero. We observed that a perfect transition
occurs without reflection under resonance, which satisfies Eq.~(\ref{E27}).
Under non-resonance, a reflected wave appears after scattering. This result
can be used to detect the eigenenergy for $H_{\mathrm{c}}$ by switching off
all leads of a multichannel scattered system (Fig.~\ref{fig1}) except $0$
and $1$, which is essentially an infinite chain sharing a single site with
the scattering center. Because perfect transmission ($r=0$) corresponds to
on-resonance, resonant energy can be determined by scanning $\mu $ and
measuring the reflection. The scattering center can be Hermitian or
non-Hermitian, but the corresponding eigenenergy must be real. The coupling
strength $J$ of the two leads is not limited; this differs from the
requirement that the coupling strength be sufficiently low for the
measurement of the eigenfunction through multichannel scattering.

To determine the validity of the method of detecting eigenenergy, we used an
SSH chain and non-Hermitian SSH chain as scattering centers to measure the
reflection of different $\mu $ values. The numerical results for several
typical sets of parameters are plotted in Fig.~\ref{fig7}(b)-~\ref{fig7}(f).
A comparison between the exact eigenenergy and the numerical results
indicated, we find that all the $\mu $ values with zero reflection
corresponded to eigenenergy, which suggested that our scattering method for
detecting eigenenergy was applicable to both Hermitian and non-Hermitian
clusters with real values. For the non-Hermitian SSH model, the
eigenenergy obtained by numerical calculation corresponds to the analytic
expression Eq.~(\ref{E23}). Therefore, we can experimentally detect the
corresponding eigenenergy by determining whether the reflection is zero in
the proposed system in Fig.~\ref{fig7}(a).

\section*{Appendix B: Analytical solutions of the transmission reflection in
SSH model}

\label{Appendix B}

The transmission coefficients of each channel are the nonnormalized
eigenfunction of the edge state. The transmission coefficients of two
adjacent odd transmission channels are $-q $ times different and all
transmission coefficients of even transmission channels are zero,
\begin{equation}
t_{l}=\left\{
\begin{array}{cc}
t_1(-q)^{(j-1)/2}, & \text{odd } \\
0. & \text{even }%
\end{array}%
\right.  \label{E29}
\end{equation}%
We supposed that the number of unit cells in the scattering center was
sufficiently large. This would guarantee that the eigenenergy of the in-gap
eigenmode would be zero, thus satisfying the resonance condition, and that
the transmission coefficients would converge, enabling the use of the
infinite series summation formula,
\begin{equation}
\sum\limits_{l=1}^{N}\left\vert t_{l}\right\vert ^{2}=\left\vert
t_{1}\right\vert ^{2}(1+q^2+q^4+\cdots)=\frac{\left\vert t_{1}\right\vert
^{2}}{1-q^{2}}.  \label{E30}
\end{equation}
Because the Hamiltonian of the entire scattering system is Hermitian, the
unitary scattering and probability current is conserved,
\begin{equation}
\left\vert r\right\vert ^{2}+\sum\limits_{l=1}^{N}\left\vert
t_{l}\right\vert ^{2}=1.  \label{E31}
\end{equation}%
The continuity condition Eq.~(\ref{E6}) and Eq.~(\ref{E30}) can be
introduced into Eq.~(\ref{E31}), we can solve that
\begin{equation}
t_{1}=\frac{2(1-q^{2})}{2-q^{2}}, r=-\frac{q^{2}}{2-q^{2}}.  \label{E32}
\end{equation}%
In the numerical simulations, the probability summations in each channel $%
p_l $ reflect reflection $\left\vert r\right\vert ^{2}$ and transmission $%
\left\vert t_l\right\vert ^{2}$. Therefore, analytical solutions for
transmission reflection can be obtained, which is shown in Eq.~(\ref{E14}).


\begin{thebibliography}{99}
\bibitem{HalperinPRB82} B. I. Halperin, Quantized Hall conductance,
current-carrying edge states, and the existence of extended states in a
two-dimensional disordered potential, Phys. Rev. B \textbf{25}, 2185 (1982).

\bibitem{CooperRMP19} N. R. Cooper, J. Dalibard, and I. B. Spielman,
Topological bands for ultracold atoms, Rev. Mod. Phys. \textbf{91}, 015005
(2019).

\bibitem{BansilRMP16} A. Bansil, H. Lin, and T. Das, Colloquium: Topological
insulators, Rev. Mod. Phys. \textbf{88}, 021004 (2016).

\bibitem{AsbothBook16} J. K. Asb\'{o}th, L. Oroszl\'{a}ny, and A. P\'{a}lyi,
\emph{A Short Course on Topological Insulators: Band-structure topology and
edge states in one and two dimensions} (Springer International Press 2016)

\bibitem{OzawaRMP19} T. Ozawa, H.M. Price, A. Amo, N. Goldman, M. Hafezi, L.
Lu, M. C. Rechtsman, D. Schuster, J. Simon, O. Zilberberg, and I. Carusotto,
Topological photonics, Rev. Mod. Phys. \textbf{91}, 015006 (2019).

\bibitem{HaldanePRL08} F. D. M. Haldane and S. Raghu, Possible Realization
of Directional Optical Waveguides in Photonic Crystals with Broken
Time-Reversal Symmetry, Phys. Rev. Lett. \textbf{100}, 013904 (2008).

\bibitem{RaghuPRA78} S. Raghu and F. D. M. Haldane, Analogs of Quantum-Hall-
Effect Edge States in Photonic Crystals, Phys. Rev. A \textbf{78}, 033834
(2008).

\bibitem{ZWangPRL08} Z.Wang, Y. D. Chong, J. D. Joannopoulos, and M. Solja%
\v{c}i\'{c}, Reflection-Free One-Way Edge Modes in a Gyromagnetic Photonic
Crystal, Phys. Rev. Lett. \textbf{100}, 013905 (2008).

\bibitem{ZWangNAT09} Z.Wang, Y. D. Chong, J. D. Joannopoulos, and M. Solja%
\v{c}i\'{c}, Observation of Unidirectional Backscattering-Immune Topological
Electromagnetic States, Nature (London) \textbf{461}, 772 (2009).

\bibitem{WJChenNAT14} W.-J. Chen, S.-J. Jiang, X.-D. Chen, J.-W. Dong, and
C. T. Chan, Experimental Realization of Photonic Topological Insulator in a
Uniaxial Metacrystal Waveguide, Nat Commun \textbf{5}, 5782 (2014).

\bibitem{RechtsmanNAT13} M. C. Rechtsman, J. M. Zeuner, Y. Plotnik, Y.
Lumer, D. Podolsky, F. Dreisow, S. Nolte, M. Segev, and A. Szameit, Photonic
Floquet Topological Insulators, Nature (London) \textbf{496}, 196 (2013).

\bibitem{HafeziNAT11} M. Hafezi, E. A. Demler, M. D. Lukin, and J. M.
Taylor, Robust Optical Delay Lines with Topological Protection, Nat. Phys.
\textbf{7}, 907 (2011).

\bibitem{HafeziNAT13} M. Hafezi, S. Mittal, J. Fan, A. Migdall, and J. M.
Taylor, Imaging topological edge states in silicon photonics, Nat. Photon., \textbf{7}, 1001 (2013).

\bibitem{KFangNAT12} K. Fang, Z. Yu, and S. Fan, Realizing Effective
Magnetic Field for Photons by Controlling the Phase of Dynamic Modulation,
Nat. Photon. \textbf{6}, 782 (2012).

\bibitem{KochPRA10} J. Koch, A. A. Houck, K. L. Hur, and S. M. Girvin, Time-
Reversal-Symmetry Breaking in Circuit-QED-Based Photon Lattices, Phys. Rev.
A \textbf{82}, 043811 (2010).

\bibitem{PetrescuPRA12} A. Petrescu, A. A. Houck, and K. Le Hur, Anomalous
Hall Effects of Light and Chiral Edge Modes on the Kagom\'{e} Lattice, Phys.
Rev. A \textbf{86}, 053804 (2012).

\bibitem{KhanikaevNAT13} A. B. Khanikaev, S. H. Mousavi, W.-K. Tse, M.
Kargarian, A. H. MacDonald, and G. Shvets, Photonic Topological Insulators,
Nat. Mater. \textbf{12}, 233 (2013).

\bibitem{SatoPRB09} M. Sato, Topological properties of spin-triplet
superconductors and Fermi surface topology in the normal state, Phys. Rev. B
\textbf{79}, 214526 (2009).

\bibitem{PesinNAT10} D. Pesin, L. Balents, Mott physics and band topology in
materials with strong Spin-orbit interaction, Nat. Phys. \textbf{6}, 376
(2010).

\bibitem{XWanPRB11} X. Wan, A. M. Turner, A. Vishwanath, and S. Y. Savrasov,
Topological Semimetal and Fermi-Arc Surface States in the Electronic
Structure of Pyrochlore Iridates, Phys. Rev. B \textbf{83}, 205101 (2011).

\bibitem{WrayNAT11} L. A. Wray, S.-Y. Xu, Y. Xia, D. Hsieh, A. V. Fedorov,
Y. S. Hor, R. J. Cava, A. Bansil, H. Lin, and M. Z. Hasan, A Topological
Insulator Surface under Strong Coulomb, Magnetic and Disorder Perturbations,
Nat. Phys. \textbf{7}, 32 (2011).

\bibitem{SYXuSCI11} S.-Y. Xu, Y. Xia, L. A. Wray, S. Jia, F. Meier, J. H.
Dil, J. Osterwalder, B. Slomski, A. Bansil, H. Lin, R. J. Cava, M. Z. Hasan,
Topological Phase Transition and Texture Inversion in a Tunable Topological
Insulator, Science \textbf{332}, 560 (2011).

\bibitem{LWuNAT13} L. Wu, M. Brahlek, R. V. Aguilar, A. V. Stier, C. M.
Morris, Y. Lubashevsky, L. S. Bilbro, N. Bansal, S. Oh, and N. P. Armitage,
A Sudden Collapse in the Transport Lifetime across the Topological Phase
Transition in (Bi$_{1-x}$In$_{x}$)$_{2}$Se$_{3}$, Nat. Phys. \textbf{9}, 410
(2013).

\bibitem{KitagawaPRB10} T. Kitagawa, E. Berg, M. Rudner, and E. Demler,
Topological characterization of periodically driven quantum systems, Phys.
Rev. B \textbf{82}, 235114 (2010).

\bibitem{ZGuPRL11} Z. Gu, H. A. Fertig, D. P. Arovas, and A. Auerbach,
Floquet Spectrum and Transport through an Irradiated Graphene Ribbon, Phys.
Rev. Lett. \textbf{107}, 216601 (2011).

\bibitem{LindnerNAT11} N. H. Lindner, G. Refael, and V. Galitski, Floquet
Topological Insulator in Semiconductor Quantum Wells, Nat. Phys. \textbf{7},
490 (2011).

\bibitem{DoraPRL12} B. D\'{o}ra, J. Cayssol, F. Simon, and R. Moessner,
Optically Engineering the Topological Properties of a Spin Hall Insulator,
Phys. Rev. Lett. \textbf{108}, 056602 (2012).

\bibitem{KatanPRL13} Y. T. Katan, and D. Podolsky, Modulated Floquet
Topological Insulators, Phys. Rev. Lett. \textbf{110}, 016802 (2013).

\bibitem{KunduPRL13} A. Kundu, and B. Seradjeh, Transport Signatures of
Floquet Majorana Fermions in Driven Topological Superconductors, Phys. Rev.
Lett. \textbf{111}, 136402 (2013).

\bibitem{YHWangSCI13} Y. H. Wang, H. Steinberg, P. Jarillo-Herrero, and N.
Gedik, Observation of Floquet-Bloch States on the Surface of a Topological
Insulator, Science \textbf{342}, 453 (2013).

\bibitem{PiskunowPRB14} P. M. Perez-Piskunow, G. Usaj, C. A. Balseiro, and
L. E. F. F. Torres, Floquet Chiral Edge States in Graphene, Phys. Rev. B
\textbf{89}, 121401 (2014).

\bibitem{RWangEPL14} R. Wang, B. Wang, R. Shen, L. Sheng, and D. Y. Xing,
Floquet Weyl Semimetal Induced by off-Resonant Light, Europhys. Lett.
\textbf{105}, 17004 (2014).

\bibitem{BernevigBook13} B. Bernevig, and T. Hughes, \emph{Topological
Insulators and Topological Superconductors} (Princeton Univ. Press, 2013).

\bibitem{OzawaPRL14} T. Ozawa, and I. Carusotto, Anomalous and quantum Hall
effects in lossy photonic lattices, Phys. Rev. Lett. \textbf{112}, 133902
(2014).

\bibitem{HafeziPRL14} M. Hafezi, Measuring topological invariants in
photonic systems, Phys. Rev. Lett. \textbf{112}, 210405 (2014).

\bibitem{WHuPRX15} W. Hu, J. C. Pillay, K. Wu, M. Pasek, P. P. Shum, and Y.
D. Chong, Measurement of a topological edge invariant in a microwave
network, Phys. Rev. X \textbf{5}, 011012 (2015).

\bibitem{AtalaNAT14} M. Atala, M. Aidelsburger, J. T. Barreiro, D. Abanin,
T. Kitagawa, E. Demler, and I. Bloch, Direct measurement of the Zak phase in
topological Bloch bands, Nat. Phys. \textbf{9}, 795 (2014).

\bibitem{DucaSCI15} L. Duca, T. Li, M. Reitter, I. Bloch, M. Schleier-Smith,
U. Schneider, An Aharonov-Bohm interferometer for determining Bloch band
topology, Science \textbf{347}, 288 (2015).

\bibitem{AidelsburgerSCI15} M. Aidelsburger, M. Lohse, C. Schweizer, M.
Atala, J. T. Barreiro, S. Nascimb\`{e}ne, N. R. Cooper, I. Bloch, and N.
Goldman, Measuring the Chern number of Hofstadter bands with ultracold
bosonic atoms, Nat. Phys \textbf{11}, 162 (2015).

\bibitem{MittalNAT16} S. Mittal, S. Ganeshan, J. Fan, A. Vaezi, and M.
Hafezi, Measurement of topological invariants in a 2D photonic system,
Nat. Photonics \textbf{10}, 180 (2016).

\bibitem{NejadPRL19} F. Z.-Nejad and R. Fleury, Topological Fano Resonances,
Phys. Rev. Lett. \textbf{122}, 014301 (2019).

\bibitem{ReisnerPRA20} M. Reisner, D. H. Jeon, C. Schindler, H. Schomerus,
F. Mortessagne, U. Kuhl and T. Kottos, Self-Shielded Topological Receiver
Protectors, Phys. Rev. Applied \textbf{13}, 034067 (2020).

\bibitem{MeidanPRB11} D. Meidan, T. Micklitz, and P. W. Brouwer, Topological
classification of adiabatic processes, Phys. Rev. B \textbf{84}, 195410
(2011).

\bibitem{FulgaPRB12} I. C. Fulga, F. Hassler, and A. R. Akhmerov, Scattering
theory of topological insulators and superconductors, Phys. Rev. B \textbf{85%
}, 165409 (2012).

\bibitem{RudnerPRX13} M. S. Rudner, N. H. Lindner, E. Berg, and M. Levin,
Anomalous Edge States and the Bulk-Edge Correspondence for Periodically
Driven Two-Dimensional Systems, Phys. Rev. X \textbf{3}, 031005 (2013).

\bibitem{PasekPRB14} M. Pasek and Y. D. Chong, Network models of photonic
Floquet topological insulators, Phys. Rev. B \textbf{89}, 075113 (2014).

\bibitem{LJinPRA17} L. Jin, Topological phases and edge states in a
non-Hermitian trimerized optical lattice, Phys. Rev. A \textbf{96}, 032103
(2017).

\bibitem{HCWuPRB19} H. C. Wu, L. Jin, and Z. Song, Inversion symmetric
non-Hermitian Chern insulator, Phys. Rev. B \textbf{100}, 155117 (2019).

\bibitem{MXiaoPRX14} M. Xiao, Z. Q. Zhang, and C. T. Chan, Surface Impedance
and Bulk Band Geometric Phases in One-Dimensional Systems, Phys. Rev. X
\textbf{4}, 021017 (2014).

\bibitem{PoshakinskiyPRA15} A. V. Poshakinskiy, A. N. Poddubny, and M.
Hafezi, Phase spectroscopy of topological invariants in photonic crystals,
Phys. Rev. A \textbf{91}, 043830 (2015).

\bibitem{ArkinstallPRB17} J. Arkinstall, M. H. Teimourpour, L. Feng, R.
El-Ganainy, and H. Schomerus, Topological tight-binding models from
nontrivial square roots, Phys. Rev. B \textbf{95}, 165109 (2017).

\bibitem{KLZhangPRB19} K. L. Zhang, H. C. Wu, L. Jin, and Z. Song,
Topological Phase Transition Independent of System Non-Hermiticity, Phys.
Rev. B \textbf{100}, 045141 (2019).

\bibitem{HCWuPRB20} H. C. Wu, X. M. Yang, L. Jin, and Z. Song, Untying links
through anti-parity-time-symmetric coupling, Phys. Rev. B \textbf{102},
161101(R) (2020).

\bibitem{HCWuPRB21} H. C. Wu, L. Jin, and Z. Song, Topology of an
anti-parity-time symmetric non-Hermitian Su-Schirieffer-Heeger model, Phys.
Rev. B \textbf{103}, 235110 (2021).

\bibitem{BYXiePRB18} B. Y. Xie, H. F. Wang, H. X. Wang, X. Y. Zhu, J. H.
Jiang, M. H. Lu, and Y. F. Chen, Second-order photonic topological insulator
with corner states, Phys. Rev. B \textbf{98}, 205147 (2018).

\bibitem{BomantaraPRB19} R. W. Bomantara, L. W. Zhou, J. X. Pan, and J. B.
Gong, Coupled-wire construction of static and Floquet second-order
topological insulators, Phys. Rev. B \textbf{99}, 045441 (2019).

\bibitem{FukuiPRB19} T. Fukui, Dirac fermion model associated with a
second-order topological insulator, Phys. Rev. B \textbf{99}, 165129 (2019).

\bibitem{LibertoPRA16} M. D. Liberto, A. Recati, I. Carusotto and C.
Menotti, Two-body physics in the Su-Schrieffer-Heeger model, Phys. Rev. A
\textbf{94}, 062704 (2016).

\bibitem{LibertoEPJ17} M. D. Liberto, A. Recati, I. Carusotto and C.
Menotti, Two-body bound and edge states in the extended SSH Bose-Hubbard
model, Eur. Phys. J. Spec. Top. \textbf{226}, 2751 (2017).

\bibitem{LLuNAT14} L. Lu, J. D. Joannopoulos, and M. Solja\v{c}i\'{c},
Topological photonics, Nat. Photon. \textbf{8}, 821 (2014).

\bibitem{RechtsmanPRL13} M. C. Rechtsman, Y. Plotnik, J. M. Zeuner, D. Song,
Z. Chen, A. Szameit, and M. Segev, Topological Creation and Destruction of
Edge States in Photonic Graphene, Phys. Rev. Lett. \textbf{111}, 103901
(2013).

\bibitem{PlotnikNAT14} Y. Plotnik, M. C. Rechtsman, D. Song, M. Heinrich, J.
M. Zeuner, S. Nolte, Y. Lumer, N. Malkova, J. Xu, A. Szameit, Z. Chen, and
M. Segev, Observation of unconventional edge states in `photonic graphene',
Nat. Mater. \textbf{13}, 57 (2014).

\bibitem{NohNAT17} J. Noh, S. Huang, D. Leykam, Y. D. Chong, K. P. Chen, and M. C. Rechtsman, Experimental observation of optical Weyl points and Fermi arc-like surface states, Nat. Phys. \textbf{13}, 611 (2017).

\bibitem{StutzerNAT18} S. St\"{u}tzer, Y. Plotnik, Y. Lumer, P. Titum, N. H.
Lindner, M. Segev, M. C. Rechtsman, and A. Szameit, Photonic topological
Anderson insulators, Nature \textbf{560}, 461 (2018).

\bibitem{NohNAT18} J. Noh, W. A. Benalcazar, S. Huang, M. J. Collins, K. P.
Chen, T. L. Hughes, and M. C. Rechtsman, Topological protection of photonic
mid-gap defect modes, Nat. Photon. \textbf{12}, 408 (2018).

\bibitem{HausIEEE91} H. A. Haus and W. Huang, Coupled-mode theory,
Proceedings of the IEEE, \textbf{79}, 1505 (1991).

\bibitem{SHFanJOS03} S. Fan, W. Suh, and J. D. Joannopoulos, Temporal
coupled-mode theory for the Fano resonance in optical resonators, J. Opt.
Soc. Am. A \textbf{20}, 569 (2003).

\bibitem{JoannopoulosBook08} J. D. Joannopoulos, S. G. Johnson, J. N. Winn,
and R. D. Meade, \emph{Photonic Crystals: Modeling the Flow of Light}
(Princeton University Press, Princeton, NJ, 2008).

\bibitem{KLZhangPRA18} K. L. Zhang, P. Wang, G. Zhang, and Z. Song, Simple
harmonic oscillation in a non-Hermitian Su-Schrieffer-Heeger chain at the
exceptional point, Phys. Rev. A \textbf{98}, 022128 (2018).
\end{thebibliography}
\end{document}